\documentclass[twocolumn,showpacs,preprintnumbers,amsmath,amssymb]{revtex4}
\usepackage[dvips]{graphicx}
\usepackage{latexsym}
\begin{document}
 
\title{Structural modulations in Sr$_{14}$Cu$_{24}$O$_{41}$
and their relation to charge ordering}

\author{M. v. Zimmermann$^1$, J. Geck$^2$, S. Kiele$^{1,2}$, R. Klingeler$^2$, and B. B\"uchner$^2$}

\affiliation{$^1$ Hamburger Synchrotronstrahlungslabor HASYLAB at Deutsches Elektronen-Synchrotron DESY,
Notkestrasse 85, 22603 Hamburg, Germany\\
$^2$ Leibniz Institute for Solid State and Materials Research Dresden at IFW Dresden, Helmholtzstr. 20, 01069 Dresden, Germany}

\date{\today}

\begin{abstract}
Structural properties of the spin chain and ladder compound Sr$_{14}$Cu$_{24}$O$_{41}$ have been studied
using diffraction with hard x-rays. 
Strong incommensurate modulation reflections are observed due to the lattice mismatch of the chain and ladder
structure, respectively. While modulation reflections of low orders display only a weak temperature independence, 
higher orders dramatically increase in intensity when cooling the sample to 10~K. All observed modulation 
reflections are indexed within the super space group symmetry and no structural phase transition could be identified 
between 10~K and room temperature. 
We argue that these modulation reflections are not caused by a five-fold periodicity of the chain lattice,
as claimed by Fukuda {\it et al.} Phys. Rev. B {\bf 66}, 012104 (2002),
but that holes localize in the potential given by the lattice modulation, which in turn gives rise to a further 
deformation of the lattice.
\end{abstract}


\maketitle

\section{Introduction}
\label{intro}

In high-temperature superconductors the correlation of spin and charge play a 
dominant role for the formation of the superconducting state. Probing the ordering
and dynamics of these degrees of freedom in materials that exhibit a different 
dimensionality might give further insights into the superconducting mechanism.  
In spin chain and ladder materials the
basic structural unit is similar to the one of high-T$_c$ superconductors. While in the 
superconductors CuO$_2$ plaquettes form a two-dimensional network, in ladder and
chain materials they compose one-dimensional arrays.
Sr$_{14-x}$Ca$_{x}$Cu$_{24}$O$_{41}$ is a family of materials that exhibits both 
CuO$_2$ chains and two-leg Cu$_2$O$_3$ ladders that are stacked in subsequent layers 
along the crystal $b$-axis. They are particular exciting since they also show 
superconductivity for x=11.5 under an external pressure of 3 GPa \cite{Uehara96},
which was theoretically predicted \cite{Sigrist94,Dagotto96}. 

The average valence state of Cu in the $x=0$ material is 2.25. 
However, optical conductivity and x-ray absorption measurements at room temperature have shown 
that out of the 6 holes per formula unit five are located on chain sites and about one hole 
resides on a ladder site \cite{Osafune97,Nucker00}. 
On the other hand more recent NMR experiments are explained in terms of a transfer 
of holes from the ladders into the chains with decreasing temperature \cite{Thurber03}.
%
The insulating character indicates that the holes are localized at low temperatures and
indeed a charge valence ordering below 200~K is claimed on the basis of results obtained by many different 
experimental techniques \cite{Ammerahl00,Takigawa98,Kataev01,Regnault99,Matsuda99,Blumberg02}.
The existence of two different Cu sites in the chains has been shown by NMR measurements
were two distinct resonances have been observed \cite{Takigawa98}.
These are assigned to two different hole sites, i.e. one located in
between two Cu spins S=1/2 forming a dimer, the other one belonging to the
holes which decouple the dimers from each other.
Above 200~K the split peak merges into a single peak due to thermal fluctuations. 
The magnetic susceptibility only slightly decreases at the charge ordering temperature of about
200 K, but dramatically decreases below 80~K indicating the formation of
spin dimers. 
By inelastic neutron scattering experiments 
a spin gap of 14 meV in the chains with a fivefold dispersion was observed \cite{Regnault99,Matsuda99}. 
A model that explains both the magnetization and the dispersion is that two Cu spins separated
by a Zhang-Rice singlet form a dimer and two dimers are separated by another two
Zhang-Rice singlets, such that the overall periodicity of the chains is five times enlarged \cite{Matsuda99}.
The hole localization also couples to the lattice as shown by thermal expansion 
experiments \cite{Ammerahl00}.
Two x-ray studies aimed to detect the structural distortion due to a hole localization
come to different conclusions. Cox {\it et al.} identify superlattice
reflections at $(0,~0,~1.25)$ and $(0,~0,~1.5)$ which are attributed to a 
quadrupling of the basic chain unit cell due to charge order. These superlattice 
reflections disappear at about 300~K. Furthermore they find a diminishing intensity of 
the $(0,~0,~2)$ Bragg reflection which is explained by a sliding of neighboring chains along 
the c-axis \cite{Cox98}. 
In contrast, Fukuda {\it et al.} report a five-fold superstructure based on the observation
of superlattice peaks at $(0,~0,~2.2)$, $(0,~0,~3.8)$ and $(0,~0,~4.8)$. These peaks show only a 
weak temperature dependence and are stable up to high temperatures \cite{Fukuda02}.
In both investigations the superlattice reflections have been ascribed solely to the 
structural properties of the chains. 
%

We have characterized a single crystal of Sr$_{14}$Cu$_{24}$O$_{41}$ with high 
energy x-ray diffraction. At room temperature we find a strong incommensurate 
modulation of both chains and ladders due to their different $c$-axis lattice
parameters. Decreasing the temperature to 10~K, additional reflections
appear that are well described within the super space group symmetry. 
We identify a continuous transition of the modulation shape from a sinusoidal 
shape to a modulation that involves various higher order Fourier components.
The origin for the variation of the distortion shape at low temperatures is
attributed to charge localization.

\section{Experimental details}
\label{Experimental}
The sample was grown by the traveling solvent floating zone
method \cite{Ammerahl98,Ammerahl99}. A piece of $3 \times 0.5 \times 2 \ \textrm{mm}^3$ size
was cleaved out of the wafer and no further surface preparation was applied. 
The experiment was performed at the high energy beamline BW5 at HASYLAB in 
Hamburg \cite{Bouchard98}. The large penetration depth of the 100 keV x-ray beam into 
the sample of about 2 mm assures that only bulk properties of the sample are probed. 
A SiGe gradient crystal \cite{Keitel98} was chosen for monochromator 
and analyzer with a convoluted full width at half maximum (FWHM) of about 40'', resulting 
in a longitudinal resolution of 0.012 \AA$^{-1}$  (FWHM) at the $(0,~0,~2)$ ladder reflection.
The high perfection of the sample mosaic of 0.005$^{\circ}$ gives a transverse
resolution of $5 \times 10^{-4}$ \AA$^{-1}$. 
The sample was mounted in the $bc$-scattering plane on the cold head of a closed cycle 
cryostat with a temperature stability of better than 0.2~K.

\section{Results}
\label{Results}

\subsection{Structural considerations}
\label{struc}
%
%
\begin{figure}[t]
  \includegraphics [width=10cm]{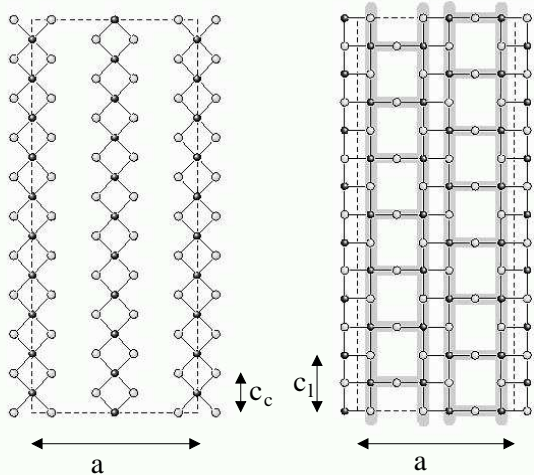}
  \caption{Sketch of the chain (left) and ladder (right) structure. The filled points
represent copper, the open circles oxygen. The Sr-layer is not shown.} 
  \label{structure}
\end{figure}
%
%
The structure of Sr$_{14}$Cu$_{24}$O$_{41}$ can be described as alternating layers
of ladders and chains, sketched in figure \ref{structure}. The ladders have a composition 
of Sr-Cu$_2$O$_3$-Sr with an orthorhombic unit cell of Fmmm symmetry and
lattice parameters of $a$=11.47 \AA, $b$=13.41 \AA{} and $c_l$=3.926 \AA. The chains are 
CuO$_2$ units with identical values of $a$ and $b$ but with $c_c$=2.744 \AA{} and 
a space group Amma \cite{McCarron88,Jensen97}.
As a result of the different lattice parameters of the chains and the ladders 
both lattices are distorted with modulated structures along their $c$-axis (see section \ref{highT} and \ref{lowT}), 
giving rise to incommensurate modulation reflections, that can not be 
indexed within the space group of either structure. However, such a modulation is described within the 
super space group formalism \cite{JannerJansen80} and the modulation reflections can be indexed by 
the four index notation $(h,~k,~l,~m)$. 
A reflection with the index
$(h,~k,~l,~0)$ is originating from the average chain structure.
Contrary, a reflection  with 
$(h,~k,~0,~m)$ stems from the average ladder structure.
Finally, a reflection with the mixed index
$(h,~k,~l,~m)$ reflects the distortion of the chain and ladder lattice due to their mutual interaction.
For a single reflection it is not possible to distinguish if it is due to a modulation of the chains
or the ladders.
%
The selection rules for the super space group Amma(001+$\gamma$)ss$\overline{1}$ with $\gamma$=0.6997(3), 
that was found for Sr$_{13.44}$Bi$_{0.56}$Cu$_{24}$O$_{41}$ \cite{Jensen97}, allows fundamental Bragg 
reflections along $(0,~0,~l)$ at $l=2n$ with n integer for the chains and $m=2n \sim l \cdot \sqrt 2$ for the ladders. 
For the $(0,~1,~l)$-scan only chain type Bragg reflections are allowed with $k+l=2n$.
Modulation peaks obey the rule $k+l+m=2n$ for both types of scans. 
To be consistent with earlier publications on x-ray diffraction on this material the index $l$ in the
figures and in the text refers to reciprocal units of the chain lattice. If a 4-index notation is used
it will be explicitly mentioned.
\subsection{Modulation reflections at T $>$ 200~K}
\label{highT}
%
\begin{figure}[t]
  \includegraphics [width=8cm]{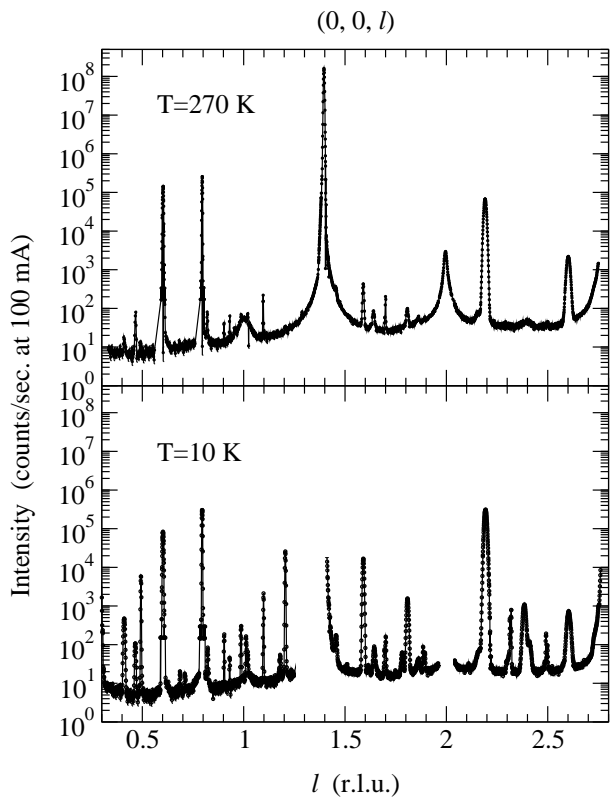}
  \caption{Scans along $(0,~0,~l)$ at 270~K and 10~K. 
The position of the fundamental Bragg reflections are at $l$=2 for the chains and 
at $l$=1.398 for the ladders and are not shown in the scan at 10~K.} \label{sl-peaks-a}
\end{figure}
\begin{figure}[t]
  \includegraphics [width=8cm]{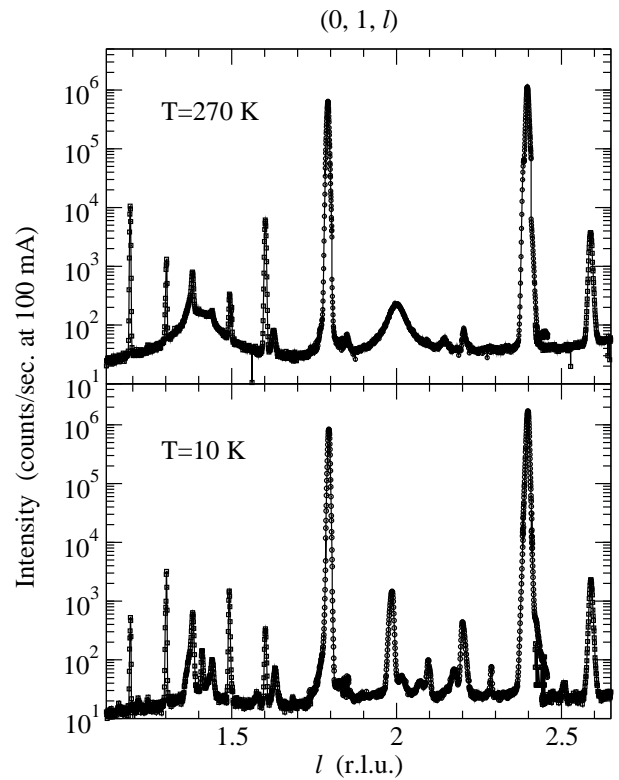}
  \caption{Scans along $(0,~1,~l)$ at 270~K and 10~K.
The position of the forbidden ladder Bragg reflection is at $l$=2.1, while the peaks
at $l$=1.8 and 2.4 are first order modulation reflections. The broad reflection at $l$=2 at
270~K is a tail of the $(0,~0,~2)$ Bragg peak (see text). At 10~K higher order reflections
appear, e.g. at $l$=2 and 2.2.} \label{sl-peaks-b}
\end{figure}
To establish the modulation of the chains and ladders we performed scans along $(0,~0,~l)$ 
and $(0,~1,~l)$ at 270~K well above the supposed charge ordering temperature. 
They are shown in figure \ref{sl-peaks-a} and \ref{sl-peaks-b}. 
The scan along $(0,~0,~l)$ shows a fundamental Bragg reflection of the chain structure at $l=2$ and for
the ladder structure at $l=1.3976(5)$ ($m=2$). Accordingly the ratio of the lattice parameters
of chains and ladders is $\gamma=c_c/c_l=0.6988(8)$, close to the value found for the Bi doped system
\cite{Jensen97}.
Additional strong reflections are observed which can be indexed using the 4-index formalism. Their 
position is listed in table~\ref{refl-positions} together with the corresponding
indices of $l$ and $m$. It can be seen that all of these strong modulation peaks observed in 
the $(0,~0,~l)$ scan are second order reflections, either a modulation of the ladders ($l$=2) or of
the chains ($m$=2). Since only second order peaks appear it is possible to distinguish between a 
distortion of the chains from a distortion of the ladders. For example, the reflection at $l$=2.601 
evidences a distortion of the chain lattice. All other reflections listed for the $(0, 0, l)$-scan in
table \ref{refl-positions} are originating from a ladder distortion. 
Higher order modulation reflections in the $(0, 0, l)$-scan are at least three orders of magnitude 
smaller in intensity.
The scan along $(0,~1,~l)$, shown in figure \ref{sl-peaks-b}, exhibits strong modulation reflections
and their positions are collected in table \ref{refl-positions}. Here modulation reflections with 
indices of $l=1$ or $l=3$ dominate the diffraction pattern. As seen in figure \ref{sl-peaks-b}, the 
first order reflections show the largest
intensity about a factor of 100 smaller than the $(0,~0,~1.3974)$ $(m=2)$ Bragg intensity. Third order reflections are 
about a factor 100 smaller than the first order reflections. (Also small reflections with $l=5$ can 
be observed.) These reflections are first and third order modulations of the ladder structure.
In general, the $(0,~0,~l)$-scan exhibits superlattice reflections of even order, while in the 
$(0,~1,~l)$-scan reflections of odd order are observed, which indicates that the phase of the modulation
of neighboring layers along the $b$-direction is shifted by $\pi$.
The small intensity of higher order reflection indicates that the modulation is of sinusoidal shape, 
leading only to second order reflections at $(0,~0,~l)$ and first and third order reflections for $(0,~1,~l)$.
%
%
Following the considerations given in section \ref{struc} and as pointed out in a recent 
comment \cite{vanSmaalen03}, it becomes clear that the primary origin of all these superlattice 
reflections is the incommensurate modulation of the chain and ladder structure due to their 
different c-axis lattice parameter and not due to a charge density wave. In particular these 
reflections are {\it not} related to a five-fold superstructure, which will be discussed in more
detail in section \ref{Discussion}.
\begin{table}[h]
\begin{tabular}{l|l}
(0, 0, l) & ( l, m) \\
\hline
0.602 & ( 2,-2)   \\
0.796 & (-2, 4)   \\
1.397 & ( 0, 2)   \\
2.000  & ( 2, 0)   \\
2.191  & (-2, 6)   \\
2.601 & ( 4,-2)   \\
3.4$^*$   & ( 2, 2)  \\
3.6$^*$   & (-2, 8)  \\ 
4.8$^*$   & ( 2, 4)  \\ 
\end{tabular}
 \hspace{1.5cm}
\begin{tabular}{l|l}
(0, 1, l)  &  ( l, m) \\
\hline
0.398$^*$  & (-1, 2) \\
1.192   & (-3, 6) \\
1.602   & ( 3,-2) \\
1.795   & (-1, 4) \\
2.400   & ( 1, 2) \\
2.588    & (-3, 8) \\
3.2$^*$ & (-1, 6) \\
3.8$^*$ & ( 1, 4) \\ 
5.2$^*$ & ( 1, 6) \\ 
\end{tabular}
\caption{Observed reflection positions and 4-index  $l$ and $m$ of the scans along $(0,~0,~l)$ 
and $(0,~1,~l)$ at 270~K. The reflections marked by a $*$ are not shown in figures \ref{sl-peaks-a}
and \ref{sl-peaks-b}.}
\label{refl-positions}
\end{table}
\begin{table}[h]
\begin{tabular}{l|l}
(0, 0, l) & ( l, m) \\
\hline
0.410  & ( 6,-8)   \\
0.493  & (-3, 5)   \\
0.987  & (-6,10)   \\
1.012  & ( 8,-10)   \\
1.098  & (-1,  3)   \\
1.205  & ( 4,-4)   \\
1.590  & (-4, 8)   \\
1.808  & ( 6,-6)   \\
2.385  & (-6,12)   \\
\end{tabular}
 \hspace{1.5cm}
\begin{tabular}{l|l}
(0, 1, l) & ( l, m) \\
\hline
0.890$^*$ & (-4, 7)   \\
1.110$^*$  & ( 6,-7)   \\
1.302  & ( 2,-1)   \\
1.410  & ( 7,-8)   \\
1.493  & (-2, 5)   \\
1.986  & (-5,10)   \\
2.093  & ( 0, 3)   \\
2.205  & ( 5,-4)   \\
2.701$^*$ & ( 2, 1)   \\
\end{tabular}
\caption{Reflection position observed at 10~K of reflections not present or of very small
intensity at 270~K.}
\label{refl-positions-10K}
\end{table}
%
%
\subsection{Modulation reflections at T $<$ 200~K}
\label{lowT}
%
\begin{figure}[t]
  \includegraphics[width=8cm]{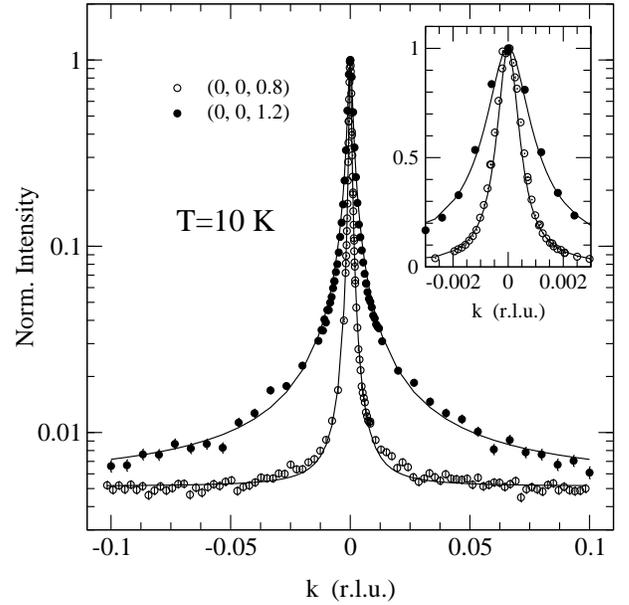}    
\caption{Reflection profiles along the $k$-direction for the modulation peaks $(0,~0,~0.8)$ and $(0,~0,~1.2)$ on 
a logarithmic scale at 10 K. The peak intensity is scaled to one for both reflections. 
The lines are fits to a Lorentzian function for the $(0,~0,~0.8)$ reflection and to a Lorentzian raised to a 
power of 1/2 for the $(0,~0,~1.2)$ reflection. The inset shows a blow-up of the peak on a linear 
scale.} \label{width-k}
\end{figure}
\begin{figure}[t]
  \includegraphics[width=8cm]{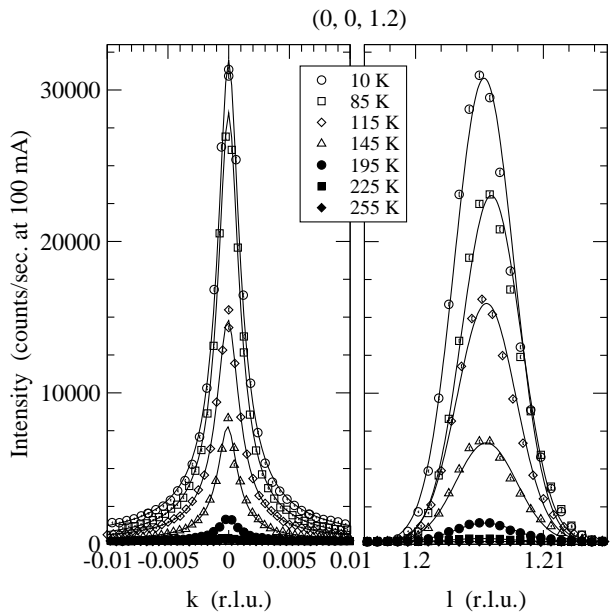}    
\caption{Scan along $k$ and $l$ at the $(0,~0,~1.205)$ reflection at selected temperature 
between 10~K and 255~K. The lines are fits to a Lorentzian raised to a power of 0.5 for 
the $k$-scan, and Gaussians for the $l$-scan.} \label{profiles-T}
\end{figure}
\begin{figure}[t]
  \includegraphics[width=8cm]{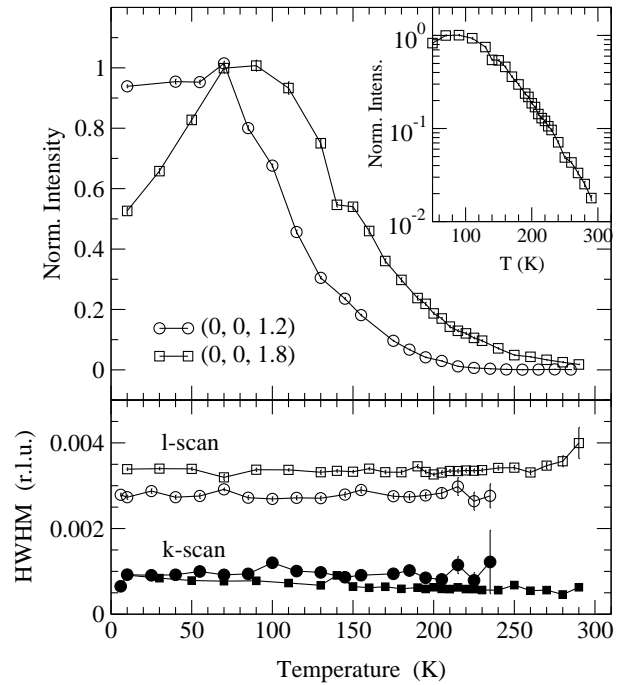}
  \caption{Top: Temperature dependence of superlattice reflections of different orders. The maximum
intensity is normalized to unity. The inset shows the intensity of the (0, 0, 1.8) reflection on
a logarithmic scale. No anomaly is visible around 200 K. 
Bottom: Half width at half maximum along $k$ (filled symbols) and $l$ (open symbols).} \label{tempd}
\end{figure}
In order to study additional distortions due to a possible charge density wave, we cooled the 
sample down to 10~K and repeated the scans along $(0,~0,~l)$ and $(0,~1,~l)$, as shown in the lower parts of figures
\ref{sl-peaks-a} and \ref{sl-peaks-b}. 
%
It is readily observed that additional superlattice reflections appear in both scans. 
The analysis of the position of all these additional reflections shows that they also belong
to the super space group symmetry, since they can be indexed within the 4-index formalism.
These additional reflections can be regarded as higher order reflections of the 
modulation observed at room temperature. We emphasize that these additional reflections do not break 
the super space symmetry.
%
In contrast to the distortion above 200 K, modulation reflections of many different orders are 
found and, consequently, it is impossible to determine whether the additional distortion at 
low temperature takes place in the chains or in the ladders.
%
In figure \ref{width-k} we compare the transverse reflection profile along $k$ of a temperature independent
reflection at $(0,~0,~0.8)$ with a reflection that shows a strong dependence on temperature, the $(0,~0,~1.2)$.
The width of the second order reflection at $(0,~0,~0.8)$ is resolution limited, and the lineshape 
is very well described by a Lorentzian scattering function
\begin{equation}
S(k)=\frac{A}{\left(1-\left(\frac{k}{\Gamma_b}\right)^2\right)^y},
\label{Lor_y}
\end{equation}
with $y=1$, amplitude A and inverse correlation length $\Gamma_b$ along the $b$-direction. The correlation length is then 
defined as $b/2 \pi \Gamma_b$. In contrast, the $(0,~0,~1.2)$ reflection shows a broadened width 
(see inset of figure \ref{width-k}) indicating a finite correlation length that we determined to 
about 3500 \AA{}.
In addition we find a peculiar line shape of this reflection, which is best described 
by a Lorentzian raised to a power of $y=1/2$, as shown by the solid line in figure \ref{width-k}.
Whether the presence of topological defects or the reduced dimensionality gives rise to the finite width 
and the unusual lineshape remains an open question. 
Along the $l$-direction both reflections are resolution limited. 
The same lineshapes with a broadened width along the $k$-direction was also observed at 
the $(0,~0,~1.8)$ reflection.

Turning to the temperature dependence of the modulation reflections we observe that
the intensity of the first and second order modulation reflections is almost independent 
of temperature as can be seen in the figures \ref{sl-peaks-a} and \ref{sl-peaks-b}. 
The difference in intensity between 10~K and 270~K is less than a factor
two (see also figure 3 in ref. \cite{Fukuda02}). The higher order reflections on the 
other hand exhibit a very pronounced temperature dependence, some of them by several 
orders of magnitude between 10~K and 270~K. 
As an example, the reflection profiles of the $(0,~0,~1.2)$ reflection perpendicular ($k$-scans)
and along the chain and ladder direction ($l$-scan) are shown 
at various temperatures in figure \ref{profiles-T}. These were fitted to a Lorentzian raised to the
power of $y=1/2$
and a Gaussian profile, respectively. The fit parameters are compiled in figure \ref{tempd}
together with the results of the $(0,~0,~1.8)$ reflection. The maximum intensity for
both reflections has been normalized to unity. Both reflections exhibit a strong increase of
their peak intensity below around 200~K and the peak intensity depends exponentially on the
temperature between 100~K and 300~K (see inset of figure \ref{tempd}). Consistent with the
absence of a reduction of the lattice symmetry, no well defined phase transition is observed. 
Rather a continuous cross-over into a distorted state is found. 
The detailed temperature dependence of the two reflections is very different. The $(0,~0,~1.2)$ reflection 
is only observable below 250~K and
shows a constant intensity below 80~K. In contrast, for the $(0,~0,~1.8)$ reflection some residual 
intensity is detectable up to 280~K and below 100~K the intensity starts to decrease.
%
As shown in the bottom part of figure \ref{tempd} no variation of the peak width could be observed 
in the temperature regime between 10~K and 250~K 
for the $(0,~0,~1.2)$ reflection. The $(0,~0,~1.8)$ shows a slight decrease of the longitudinal peak width
with increasing temperature, which could be an indication for the relaxation of lattice strain.
The basically constant width indicates that the length scale of the ordering pattern is temperature
independent. 
%
Neither temperature dependence could be observed in the position of these reflections, which is consistent
with the fact that these modulation reflections belong to the space group symmetry and that
their position is given by the mismatch of the lattice parameters of chains and ladders. 

\section{Discussion}
\label{Discussion}
The picture that emerges from the measurement of the modulation reflections is as follows:
At high temperatures
(270~K) both the chain and the ladder lattices are distorted with a mainly sinusoidal shape due to the mismatch
of their respective $c$-axis lattice parameters. Accordingly,
the distortion gives rise to modulation reflections of low orders, that we observe around 
both the ladder and the chain Bragg peaks. Thus, not only the chain lattice, but also the ladder
lattice is distorted at high temperatures. 
At low temperatures (10~K) the modulation deviates from the sinusoidal shape
leading to the appearance of higher order modulation peaks. We stress again that none of these 
modulations breaks the super space symmetry of the lattice. However, within each sublattice the
symmetry is reduced. 
The variation of the modulation shape is a continuous process, as seen from the temperature dependence 
of the higher order modulation reflections. No well defined transition temperature can be identified
from our structural studies. 
While the high temperature modulation with sinusoidal shape is long range ordered in all directions, the 
alteration of the shape
at low temperature is of intermediate range order of a length scale of a few thousand \AA ngstr\o m.
The anisotropy of the peak width shows that the interactions leading to a deviation from the sinusoidal shape are 
stronger within a chain or ladder than the ones among subequent chains or ladders.

It is natural to associate the origin of the temperature dependent distortion with the localization 
of the doped holes. This is supported by the fact that the kink in the temperature dependent 
resistivity curve \cite{Blumberg02} coincides with the temperature were the intensity of the higher order modulation 
reflections starts to rise steeply.
We note that the technique of hard x-ray diffraction is not sensitive to the holes itself, but
to lattice distortions caused by the hole ordering. The ordering pattern of the holes has to be 
such that the lattice symmetry is not broken. It is therefore most likely that the holes localize
in the potential given by the incommensurate lattice distortion at high temperatures above 200~K. 
The localization of charge carriers distorts the lattice further and leads to a variation of the shape of the
distortion, giving rise to the appearance of higher Fourier components. The amplitude and the periodicity
of the high temperature distortion remains mostly unaltered.
%
Our data show that the presence of superlattice reflections at $l$=2.2 and $l$=4.8 \cite{01l-refl} 
can not be taken as evidence for a five-fold modulation of the chain lattice \cite{Fukuda02}. A 
five-fold modulation would result in superlattice
reflection at both $l$=1.8 and $l$=2.2 which are both observed at low temperatures. However, as shown in 
section \ref{Results} the properties of the peak at $l$=1.8 are completely different from the ones 
of the $l$=2.2 reflection.
The latter is very strong, three orders of magnitude smaller than the ladder Bragg peak at 270 K, and the 
intensity is independent of the temperature. In contrast, the peak at $l$=1.8 is very small, eight 
orders of magnitude smaller than the ladder Bragg peak at 270 K, and the intensity shows a very 
strong temperature dependence (see figure \ref{tempd}). 
This clearly demonstrates that the interpretation of these superstructure
reflections in terms of a simple five-fold modulation is not possible.
In particular, indexing the modulation reflection 
using the super space symmetry we find that the $(0,~0,~1.8)$ reflection is a sixth
order modulation of the chain lattice and also sixth order of the ladder lattice. 
The $(0,~0,~2.2)$ reflection is a second order modulation reflection of the ladder 
lattice and similar intensities and similar temperature dependences for reflections 
of the same order are obtained. Not a single ordering wave vector, but a multitude 
of Fourier components characterizes the low temperature distortion. 
%
However, the dimer model which has already been described above may serve
as a first approximation for the complicated structural modulation which
takes place in the chain sublattice. In particular, inelastic neutron
scattering data of Sr$_{14}$Cu$_{24}$O$_{41}$ can be described in terms of the dimer model
\cite{Regnault99,Matsuda99}. Nonetheless,  we stress that the high energy
x-ray diffraction experiments unambiguously reveal a much more complicated
modulation in the chain and in the ladder sublattice. Since we argue that
this modulation generates a pinning potential for the holes in the chains,
we claim that the charge ordering pattern is more complex and may be more
adequately be described by a charge density wave.
This interpretation is further strenghtened by inelastic neutron
scattering experiments on
La$_{1}$Sr$_{13}$Cu$_{24}$O$_{41}$, which  contains half doped spin chains. Although the charge
carrier concentration in the chains differs considerably as compared to
Sr$_{14}$Cu$_{24}$O$_{41}$, a magnetic excitation corresponding to fluctuating spin dimers
has been observed in the lanthanum doped compound. This result supports our
scenario of charge localization due to a pinning potential and makes a
purely electronic origin of the charge and dimer ordered state very
unlikely.
%
From our present data it is not possible to construct an alternative model for the hole ordering.
This would require a full structural determination, were integrated intensities of many 
modulation reflections are measured. It is, however, experimentally challenging to determine these for 
the rather diffuse reflections.
%
Two features of our data are highly unusual, first the peculiar reflection profile perpendicular 
to the ladder/chain direction, that is described by a Lorentzian raised to a power of one-half, and
second the exponential dependence of the intensity of the higher order reflections on the temperature.  
The latter feature is closely related to the temperature dependent intensity of the chain Bragg 
reflection whose origin is a very soft thermal mode \cite{vZimmermann_tobe}. This indicates that 
the hole ordering might be strongly influenced by the chain lattice dynamics and vice versa.

\section{Concluding Summary}
\label{CS}
We have reported on the incommensurate lattice distortions in Sr$_{14}$Cu$_{24}$O$_{41}$ and shown
that the crystal structure of this composite system exhibits a strong mostly temperature independent 
lattice distortion due to the incommensurate ratio of the lattice parameters of the chains and the 
ladders, respectively. Upon cooling the sample to 10~K, additional
modulation reflections appear, that can be described as higher order reflections of the distortion
present at room temperature. These low temperature reflections indicate a variation of the shape of the
distortion pattern due to the localization of holes. It is found that holes localize in the lattice potential 
given by the high temperature distortion, since the periodicity of the modulation is not
affected by the charge localization.

\section{Acknowledgment}
The authors thank M. Gr\"uninger and S. van Smaalen for fruitful discussions. The technical support
by R. Novak and T. Kracht is highly appreciated. The work was supported by the DFG through SPP 1073.

\end{document}